# Implementation of a Sustainable Security Architecture using Radio Frequency Identification (RFID) Technology for Access Control.

By

[1]KASSIM, Shakiru Olajide; [2]Aisha Samaila Idriss and [3]Abdullahi Isa Ahmed

[1, 2 & 3]Department of Computer Engineering technology, The Federal polytechnic Damaturu, Yobe State.

## ABSTRACT

Implementation of a sustainable security architecture has been quite a challenging task with several technology deployed to achieve the feat. Automatic IDentification (Auto-ID) procedures exist to provide information about people, animals, goods and products in transit and found several applications in purchasing and distribution logistics, industries, manufacturing companies and material flow systems. This work focuses on the development and implementation of an access control system using Radio Frequency Identification (RFID) technology to enhance a sustainable security architecture. The system controls access into a restricted area by granting access only to authorized persons, which incorporates the RFID hardware (RFID tags and readers and their antennas) and the software. The antenna are to be configured for a read range of about 1.5 m and TMBE kit reader module was used to test the RFID tags. The encoding and decoding process for the reading and writing to the tag as well as interfacing of the hardware and software was achieved through the use of a FissaiD RFID Reader Writer. The software that controls the whole system was designed using in Java Language. The database required for saving the necessary information, staff/guest was designed using appropriate DataBase Management System (DBMS). The system designed and implemented provide records of all accesses (check-in and check-out) made into the restricted area with time records. Other than this system, Model based modeling through the MATLAB/Simulink, Arduino platform, etc. can be used for similar implementation.

## INTRODUCTION

One area which has been of great concern to the entire world is the security of life and property because of the emerging technology approaches to crime by perpetrators. Several technological approaches has been applied to curtail the menace of the security challenges using the available modern technologies. Sustaining such approaches has been an issue which are yet to address as most tends to be little or no effective with time as they continually used. The use of simple modern technologies in line with approaches of advanced modern technologies could be used to achieve greater sustainability of security architectures.

Access control to restricted facility using simple modern technology can enhance the security architecture of any facility demanding strict restriction and monitoring. Radio Frequency IDentification (RFID) technology is a simple and less expensive Automatic Identification (Auto-ID) technology that has been applied to different area for the control one process or the other with greater performance and success. RFID over the year has gained immense popularity for curbing security issues, most predominantly when used in conjunction with biometric authentication technologies. As an Auto-ID system, it is capable of transmitting data without the usage of any guided media with efficient reception. Hence, it is a relatively inexpensive technology that can be used to promote and provide security solutions (Md.Kishwar, et al., 2015).

Amidst of a wide range of Auto-ID systems, which includes magnetic stripes, Optical Character Recognition (OCR), barcodes, biometrics, contact memory buttons, and smart cards, the contactless capability of RFID makes it a more efficient technology for Auto-ID especially in areas of security and supply chain management in the rise of e-commerce globally (Amit & Hal, 2011). As reported by (Shoewu & Badejo, 2006; Amit & Hal, 2011; Md.Kishwar, et al., 2015; Peter, Joseph, & Gabriel, 2016; Orji, Oleka, & Nduanya, 2018)the RFID technologies have been found to be reasonably efficient in the provision of security and that its application can been found in the fields of business, home automation, industry and logistics support in particular due to its capability to detect, track, classify and manage the flow of information systematically.

In this work, the development and implementation of an access control system using RFID technology to enhance a sustainable security architecture was described. The system controls access into a restricted area by granting access only to authorized persons.

**REVIEWED RELATED LITERATURE**

Several access control system has been developed by various researchers on the use of the RFID technology for security purposes. The design of RFID based security and access control system for use in hostels was presented by (Umar, Mahmood, Muhammad, Athar, & Muhammad, 2014). The system combines RFID technology and biometrics to accomplish the required task. Implemented inside the Punjab University premises, the RFID reader installed at the entrance of hostel detects a number, captures the user image and scans the database for a match. The system grants access if both the card and captured image belong to a registered user, otherwise the system turns on the alarm and makes an emergency call to the security van through GSM modem. In a similar work, (Peter, Joseph, & Gabriel, 2016) uses an RFID tag which contains integrated circuit that is used for storing, processing unique information, modulating and demodulating the radio frequency signal being transmitted and a Global System for Mobile Communication (GSM) technology to communicate to security personnel via Short Message Service (SMS) in order to enhance the security of a conditioned environment.

Adebayo, et al., (2017), developed a simple security system that combines RFID and password to provide access control for entrances. The system implemented with ATMEGA8 Microcontroller directs the RFID reader to scan and authenticate users' identification tag which further request for password before activating a motor to grant access. The Microcontroller was programmed using Microbasic language; the data of identification tags and password were stored in its database. Results shows that the RFID reader can scan tags within 25cm range in two seconds, and open a prototype door within 3 seconds on keying in the correct password. The door controlled by motor closes after a preset 5 seconds delay. The system which was cost effective proves formidable and reliable as it was implemented for 12 users with greater performance.

In a similar work, Joseph & Sohail (2018) designed and implemented an RFID based access control system to gain access to a secure area in which the system reads an RFID card and unlocks a door to permit access. For increased security, the access configuration includes a keypad which requires a user to provide Personal Identification Number (PIN) which activates the locking mechanism to rotate servo which open the door. Also included was a small LCD screen to convey information to the user. Here, the overall control function for this system was provided by an Arduino UNO microcontroller chip. Using Arduino Orji, Oleka, & Nduanya (2018) designed an automatic RFID-based access control system. In the system, the RFID reader installed at an entrance detects an RFID tag which carries a unique identifier (UID) and compares it with the stored UID for a match, and grants access if it matches or otherwise denies.

As an Access Control System, this work controls and monitors accesses of individuals into restricted area, Computer Engineering Department (organization) using the RFID Technology. It also produce general reports of these accesses with attendance reports, thus controlling the inflow and outflow of visitors within the Department.

## METHODOLOGY: SYSTEM DESIGN AND IMPLEMENTATION
### The RFID Technology

The RFID is the combination of radio broadcast technology and radar, it consists of three main parts:

- *The Tag (Transponder)*: made up of an IC and an antenna. The IC consists of a memory for storing data and has the capability of processing data from radio frequency digital data and vice versa (Encoder and Decoder). The antenna is the physical interface for the RF to be received and transmitted.
- *The Reader (Interrogator)*: transmits an RF wave to the tag, it is made up of an Antenna and Transceiver. The antenna is the source of the radio signal that activates the tag and reads and writes data to it which is in turn sent to the PC for processing.
- *The Personal Computer (PC)*: used for general enterprise applications of the system, it processes the information received by reader via the interface program (high level language) developed for interfacing between them.

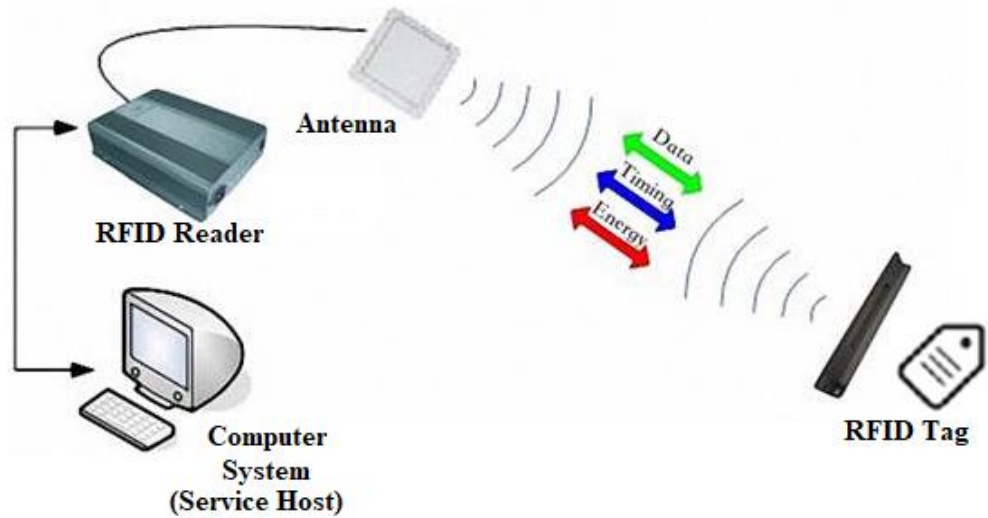

Figure 1: RFID system working principle diagram.

**System Design**

The system design is made up of two sections: the hardware and software. A general system architecture is shown in figure 2.

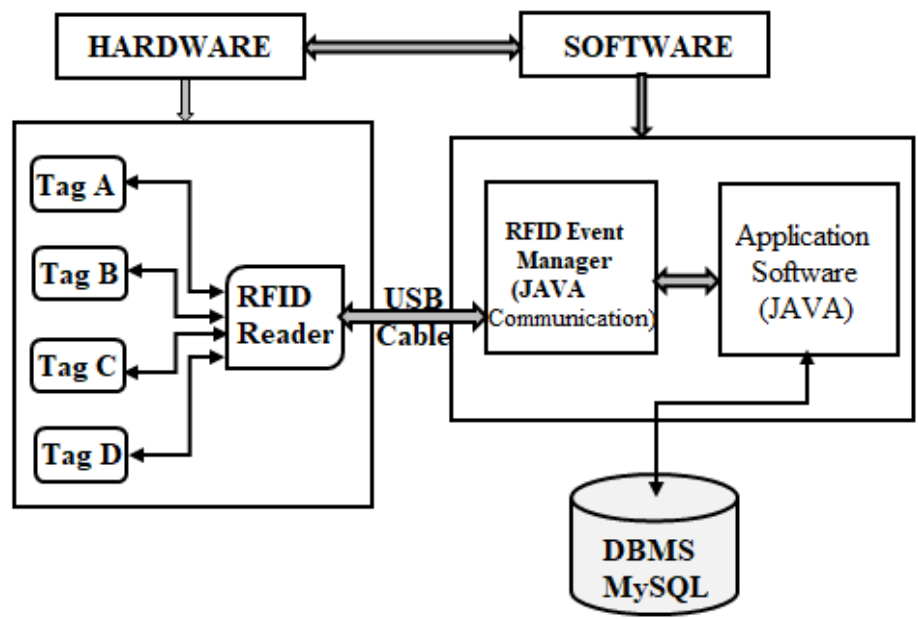

**Figure 2:** General RFID system Architecture

**Hardware design**

*Tag Selection*

In selecting the following as recommended by (Atlas RFID, 2019)were considered:
- type of surface the tagging will be used (On metal, plastic, wood, etc.);
- desired read range;
- Size limitations;
- environmental conditions to consider (excessive heat, cold, moisture, impact, etc.)

- method of attachment (adhesive, epoxy, rivets/ screws, cable ties, etc.)

Considering the system requirement (i.e. Access Control), an inlay read/write low frequency passive tags were consider for the design. The frequency selected was 125 KHz which was based on the standard stipulated by (Atlas RFID, 2019; RFID4U, 2021; Impinj, 2021), and the FID read range was about 1.5 m (near range is about 90 cm). For this work, the YARONGTECH 125 KHz rewritable T5577 Sticker Coin Adhesive Label RFID Tags selected, which has diameter of 25 mm and 1 mm thick. This can easily be affixed to the users' identification (ID) card.

*RFID Reader*

Being the brain of the RFID system, the RFID interrogators transmit and receive radio waves in order to communicate with RFID tags. They are typically divided into two distinct types: Fixed and Mobile RFID interrogators. For this pilot design, a single fixed type with built-in antenna (though has an additional antenna port for the connection of an optional external antenna) was used. This was placed at on wall of the pilot area for the test.

A FissaiD RFID Reader Writer 125 KHz (H-ID/AW-ID/EM-ID) was purchased and utilized for this work. It allows read, write and copy processing and works with the windows operating systems, though rejects other tags but was found to be compatible with the YARONGTECH 125 KHz rewritable T5577 Sticker Coin Adhesive Label RFID Tags selected.

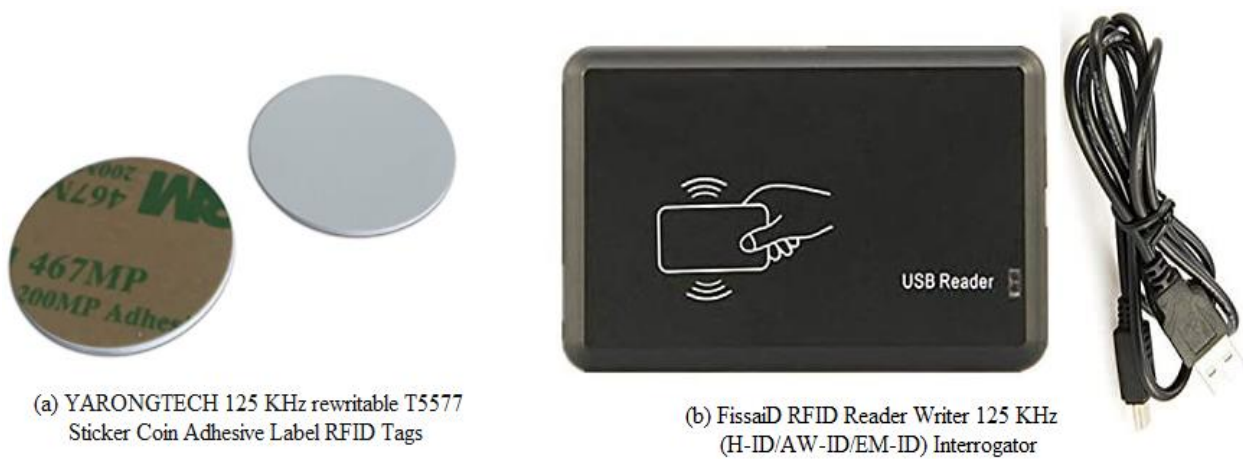

(a) YARONGTECH 125 KHz rewritable T5577 Sticker Coin Adhesive Label RFID Tags

(b) FissaiD RFID Reader Writer 125 KHz (H-ID/AW-ID/EM-ID) Interrogator

Figure 4: RFID Transponder and Interrogator

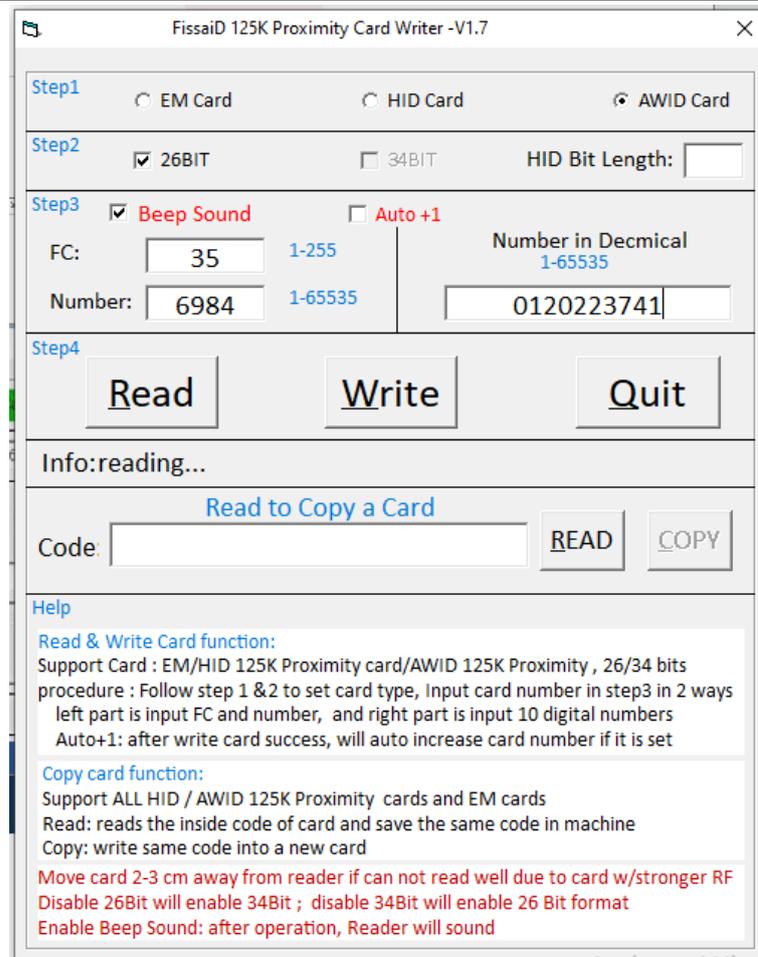

Figure 5: FissaiD RFID Reader Writer Tags Configuration window

**Software Analysis and Design**

*Event Manager: Reader/PC Interface*

Whenever a tag comes within the reader's read-range, the Event Manager serves as the monitor which notifies the Application Software. The Event Manager was developed using the JAVA(tm) Communications API Version 3.0 using the Low-level classes (SerialPort and ParallelPort) providing an interface to physical communications ports. This part of the software interacts with the PC through the serial (RS-232) and parallel (IEEE 1284) ports via a converter to the RFID interrogator USB communication cable.

The Event Manager also serves as the intermediary between the high level user and the RFID Reader when data is to be written to a tag ant it actually communicates with the controller (used for coding and decoding of data) that is embedded as part of the Reader. The data that are read (coded or decoded) are transferred to the Application Software for further correspondence with the system.

*Application* Software

The main control of the whole system is performed by the Application Software serves as the, the context that depicts the entire event of the system control is given in figure 6.

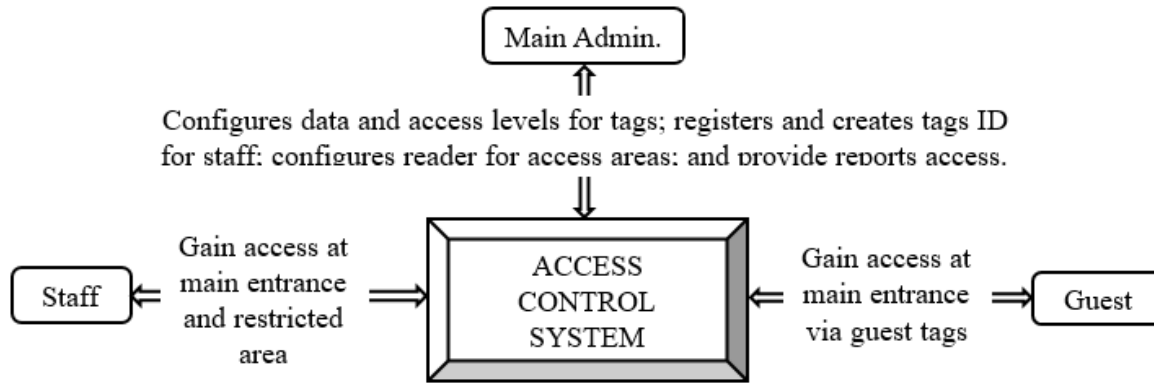

Figure 6: Application Software Event Control Flow

The application software implemented among other things performs the following:

- Configuration: defines IDs for the readers, assign the reader to restricted area and writes data (IDs) to tags. It also configures the serial port parameters and establishes communication with the RFID hardware system. It equally configures access level definitions for tags and restricted doors, creating Sections, staff registration and assign tags to staff
- Database Setup: all configurations (data) carried out by the application software are saved in the database. The setup also allows queries and processes data from the Database for verification during access requests and reports generation.
- Control: overall control of the RFID System is carried via application software such as on/off of the reader, setting the reader scan and data reading from tags, writing data to tags through the reader and general control of all the events taking place within the system.
- Report Generation: application software interacts with the database to generate reports for analysis such as report of all accesses made with the corresponding dates and times can be viewed from the application software.

Figure 7 shows the process flow for the system.

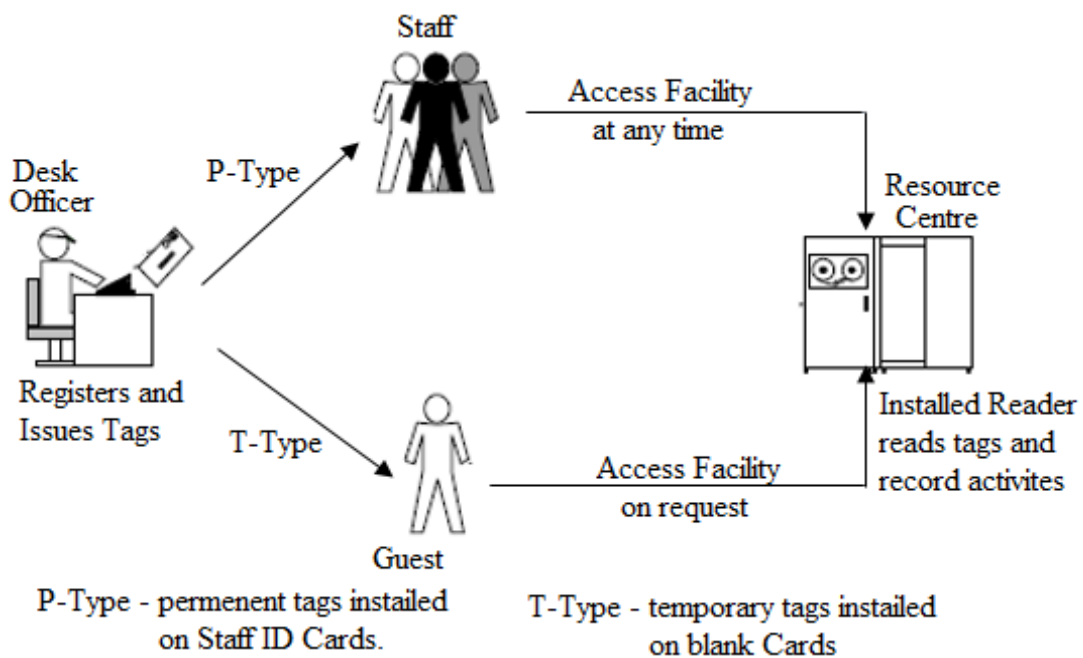

Figure 7: Process flow for the RFID Access control System

*Database Management System and its configurations*

For the DBMS, a script for the Structured Query Language (SQL) is written for the initial design of the database for the whole system. The database was fed with the necessary data made by administrator which includes all assignments, definitions and setups with respect to the tags, reader and application software.

To enable the Access Control System becomes active, the database was configured, the two major configurations are:

- Tags configuration with their respective unique IDs as mimicked in software format, i.e. tag IDs and tag type (staff and guest).
- The RFID reader unique ID which allows the Application System to know the accesses at a particular time and by who (staff or guest).

## RESULTS AND DISCUSSION

**Reader -Tag Antenna Performance Test**

Using the TMBE RFID kit which operates at same frequency (125 KHz) and similarly compatible to the design, Tag-Reader antenna response were tested. Results obtained is shown in table 1

Table 1: Results of Reader -Tag Antenna Performance Test

| Reader -Tag Antenna Distance (cm) | Reader -Tag Antenna Angle (degrees) | Tag Induced Emf (Volts) |
|---|---|---|
| 25 | 0 | 13.71 |
| 25 | 180 | 1.32 |
| 50 | 0 | 12.14 |
| 50 | 180 | 1.07 |
| 75 | 0 | 11.56 |
| 75 | 180 | 0.85 |
| 100 | 0 | 10.93 |
| 100 | 180 | 0.76 |
| 150 | 0 | 10.05 |
| 150 | 180 | 0.23 |

From the results obtained, the induced emf varies with the distance and angle between the tag antenna and the reader antenna. This indicates that the reader reads the antenna better when the angle between the tag antenna and the reader antenna is zero degree, the reception which obvious is better at short distance.

**Database System Configuration Results**

The display window of all saved data with respect to configured staff/Guest details during the configuration of the database is shown in Figure 8.

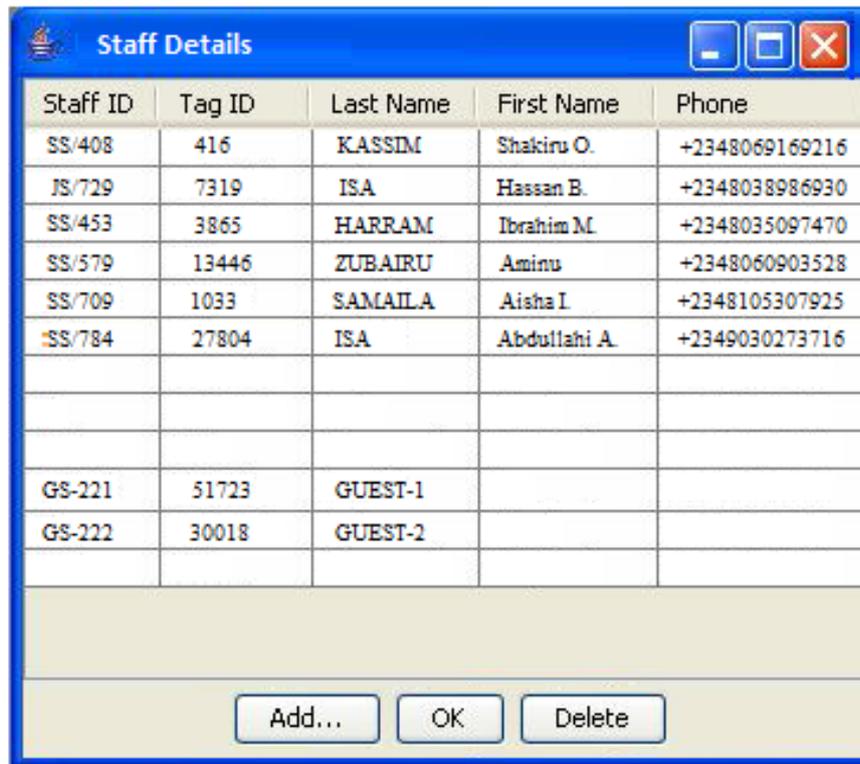

Figure 8: Database window for saved Staff Details

**System Performance Test Results**

Several tests were carried after connecting the reader using different configured staff tags affixed to their staff ID cards and guest tags affixed to a blank card. The reader was initiated to start scanning and configured staff/guest tags were brought close to the reader. For better analysis of the results returned when report is being requested, the test was carried out for two days. A status window displaying the reader connection status, its scanning status, all recent and last accesses made by the staff with tag JS/729 detected is shown in figure 9.

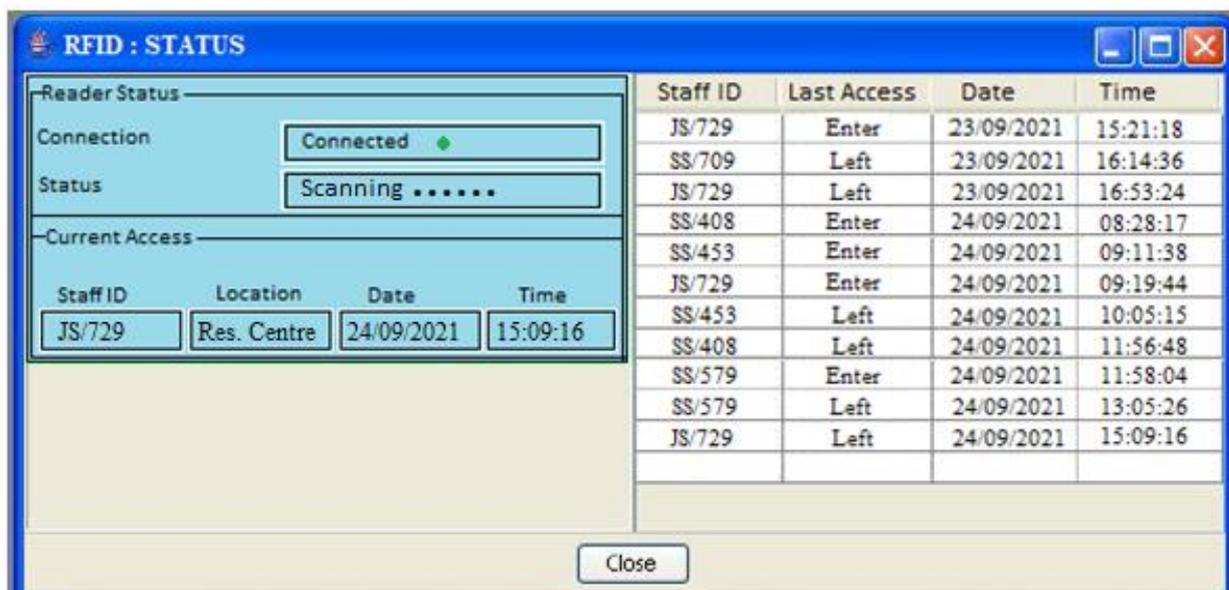

Figure 9: Window displaying the reader status and accesses

**Access Report**

The Access Report provide all accesses made by all staff/guest that visited the resource centre with their respective date and time for check-in and check-out. See figure 10 below.

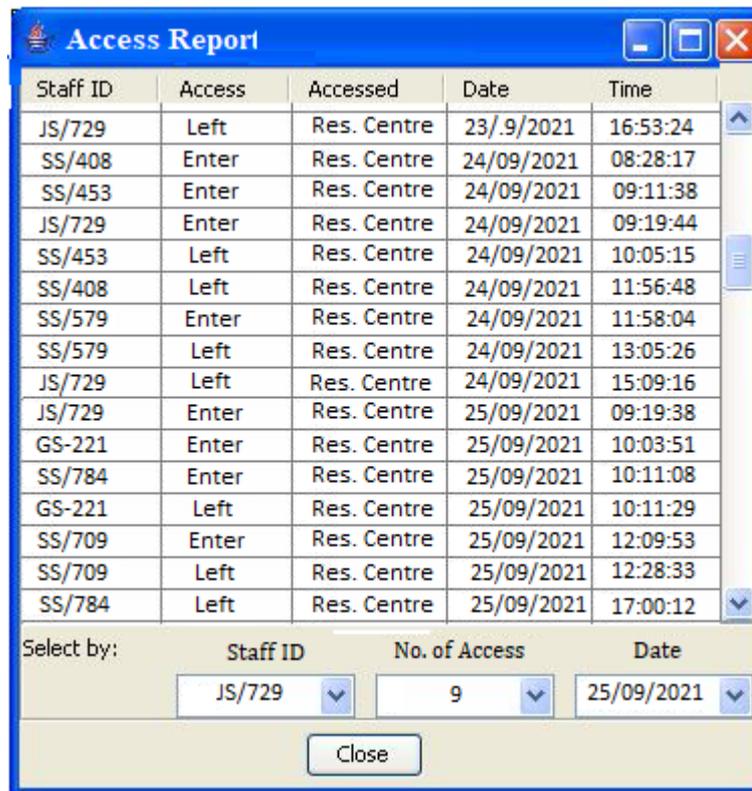

Figure 10: Window showing accesses made to resource centre by staff/guest

**Conclusion and/ Recommendation**

From the results obtained, the RFID Reader-Tag antenna test indicates that the reader can be able to scan and read the tag at a distance of up to 150 cm (i.e. 1.5 m). The overall system configuration was a success as the application software was able to provide the required configuration and assignment of unique IDs to each tags i.e. tag IDs and tag type (staff and guest). The system on being initiated proves to be active as evident in the Reader scanning and connection status, and response to tags within its vicinity. The system capability to keep records of all accesses by staff and guest to the resource centre proves that the system can actually perform the required function of access control.

The system which is invariably a model for future expansion it can be provided with other features for robustness and more functionality enhancement. Features such as attendance register for the organization, access to facilities in terms of rejection or acceptance to access a restricted facility can be incorporated. In both cases, more than one reader or more antenna in a reader can be included in the RFID system set-up.